\documentclass[onecolumn,useAMS,usenatbib]{mn2e}

\usepackage{graphicx}
\usepackage{subfigure}
\usepackage{xcolor}
\usepackage{mathtext,bbm,amsmath,amsfonts,amssymb,indentfirst,syntonly,graphicx}
\usepackage{mathtools}
\usepackage{slashbox}
\usepackage[english]{babel}
\usepackage{calc}
\usepackage{tikz}

\def\bc{\begin{center}}
\def\ec{\end{center}}
\def\be{\begin{eqnarray}}
\def\ee{\end{eqnarray}}

\title[Are long gamma-ray bursts standard candles?]{Are long gamma-ray bursts standard candles?}
\author[H.-N. Lin, X. Li, S. Wang and Z. Chang]
        {Hai-Nan Lin$^{1}$\thanks{e-mail: linhn@ihep.ac.cn.},
        Xin Li$^{1,2}$\thanks{e-mail: lixin1981@cqu.edu.cn.},
        Sai Wang$^{2}$\thanks{e-mail: wangsai@itp.ac.cn.} and
        Zhe Chang$^{3}$\thanks{e-mail: changz@ihep.ac.cn.}\\
$^{1}$Department of Physics, Chongqing University, Chongqing 401331, China\\
$^{2}$State Key Laboratory Theoretical Physics, Institute of Theoretical Physics, Chinese Academy of Sciences, Beijing 100190, China\\
$^3$Institute of High Energy Physics, Chinese Academy of Sciences, Beijing 100049, China}
\begin{document}

\date{Accepted xxxx; Received xxxx; in original form xxxx}

\pagerange{\pageref{firstpage}--\pageref{lastpage}} \pubyear{2015}

\maketitle

\label{firstpage}

\begin{abstract}
Gamma-ray bursts (GRBs) are widely proposed as an effective probe to trace the Hubble diagram of the Universe in high redshift range. However, the calibration of GRBs is not as easy as that of type-Ia supernovae (SNe Ia). Most calibrating methods at present make use one or some of the empirical luminosity correlations, e.g., Amati relation. One of the underlying assumptions of these calibrating methods is that the empirical correlation is universal over all redshifts. In this paper, we check to what extent this assumption holds. Assuming that SNe Ia exactly trace the Hubble diagram of the Universe, we re-investigate the Amati relation for low redshift ($z<1.4$) and high redshift ($z>1.4$) GRBs, respectively. It is found that the Amati relation of low-$z$ GRBs differs from that of high-$z$ GRBs at more than $3\sigma$ confidence level. This result is insensitive to cosmological models. We should be cautious when using Amati relation to reconstruct the  Hubble diagram of the Universe.
\end{abstract}


\section{Introduction}\label{sec:introduction}

Gamma-ray bursts (GRBs) are the most luminous explosions in the Universe since the big bang. The isotropic equivalent energy they released in a few seconds can be as large as $10^{48} \sim 10^{55}$ ergs. For recent reviews, see, e.g., \citet{Piran:1999,Meszaros:2002,Meszaros:2006,Kumar:2015}. Thanks to their extreme brightness, GRBs are detectable up to redshift $z\gtrsim 9$ \citep{Salvaterra:2015}. For example, the most distant GRB known today is GRB 090429B, whose redshift is as high as $z\approx 9.4$ \citep{Cucchiara:2011}. Due to their high redshift properties, GRBs are often proposed as potential candles to trace the Hubble diagram of the Universe in the high redshift range. In fact, GRBs have already been widely used, together with other candles, such as type-Ia supernovae (SNe Ia), to constrain the cosmological parameters \citep{Schaefer:2003,Bloom:2003, Xu:2005,Firmani:2005,Liang:2005,Firmani:2006a,Schaefer:2007,Liang:2008a,Liang:2008b,Wei:2009,Wei:2010,Wang:2011,Capozziello:2012,Wei:2013,Velten:2013, Cai:2013,Breton:2013,Chang:2014, Cano:2014,Cuzinatto:2014,Wang:2014,Wang:2015,Li:2015}. The consistent luminosities of SNe Ia make them the ideal distance indicators in tracing the Hubble diagram of the local (low-redshift) universe. However, since we have little knowledge about the explosion mechanism of GRBs, the GRB candle is much less standard than the SN Ia candle.

Nevertheless, one can still calibrate GRBs using the empirical luminosity correlations found in long GRBs. These correlations includes Amati relation ($E_{\rm peak}-E_{\rm iso}$) \citep{Amati:2002,Amati:2003,Amati:2006}, Ghirlanda relation ($E_{\rm peak}-E_{\gamma}$) \citep{Ghirlanda:2004b}, Yonetoku relation ($E_{\rm peak}-L_{\rm iso}$) \citep{Yonetoku:2004}, Liang-Zhang relation ($t_b-E_{\rm peak}-E_{\rm iso}$) \citep{Liang:2005}, Firmani relation ($T_{0.45}-E_{\rm peak}-L_{\rm iso}$) \citep{Firmani:2006b}, lag-luminosity relation ($\tau_{\rm lag}-L_{\rm iso}$) \citep{Norris:2000}, variability-luminosity relation ($V-L_{\rm iso}$) \citep{Fenimore:2000,Reichart:2001}, and so on. Among these luminosity correlations, the Amati relation is most widely used. This is partly because that the spectrum properties such as the peak energy $E_{\rm peak}$, the spectrum indices $\alpha$ and $\beta$, and the photon fluence $S$ which are necessary to analyze the Amati relation can be easily observed with enough precision, so that the number of available GRBs is large. Unfortunately, all of these correlations depend on a specific cosmological model. Therefore, circularity problem occurs when using GRBs to constrain the cosmological parameters. Recently, some model-independent methods were proposed to calibrate GRBs, such as the Bayesian method \citep{Firmani:2005}, the luminosity distance method \citep{Ghirlanda:2004a,Liang:2005}, and the scatter method \citep{Ghirlanda:2004a}, etc.. Actually, these methods still can't completely solve the circularity problem.

A completely model-independent method free of circularity problem is using distance ladder to calibrate GRBs \citep{Liang:2008a,Liang:2008b,Wei:2009,Wei:2010}. The main procedures are as follows: Firstly, calculate the distance moduli for the low-redshift (e.g., $z<1.4$) GRBs by using cubic interpolation from SNe Ia (e.g., Union2). Thus, the distance as well as the isotropic equivalent energy of low-$z$ GRBs can be derived. Then we can obtain the empirical luminosity correlations (such as Amati relation) from the low-$z$ GRBs. By directly extrapolating the empirical luminosity correlations to high-$z$ (e.g., $z>1.4$) GRBs, we can inversely obtain the distance moduli for high-$z$ GRBs. Since the distance moduli of SNe Ia are directly extracted from their light curves without involving any cosmological model, this calibrating method is of course completely model-independent. Recently, a similar method was proposed by \citet{Liu:2014}. The only difference is that they used the Pad\'{e} approximation \citep{Pade:1892} instead of the cubic interpolation to derive the distance moduli of low-$z$ GRBs. An underlying, but unproven assumption of these methods is that the empirical luminosity correlations are universal over all redshifts. If the empirical luminosity correlations evolve with redshift, these calibrating methods might be invalid. In fact, \citet{Wang:2011} have already investigated six empirical luminosity correlations in different redshift ranges. They found that the slope of Amati relation of high-$z$ GRBs is smaller than that of low-$z$ GRBs, although the intercept does not vary significantly with redshift. Similar features were found in the rest five luminosity correlations. Due to the large uncertainties, they concluded that no significant evidence for the redshift evolution of the luminosity correlations was found. However, in another paper \citep{Li:2007}, the author has showed that the Amati relation varies with redshifts systematically and significantly.

In this paper, we focus on checking the redshift dependence of Amati relation. Firstly, we use the Union2.1 dataset \citep{Suzuki:2012} to constrain the Hubble diagram of the Universe in the redshift range of $z<1.4$. Then we directly extrapolate the Hubble diagram to high redshift range. By assuming that GRBs follow the same Hubble diagram, we can calculate their luminosity distance and isotropic equivalent energy. Finally, we study the Amati relation for the low-$z$ ($z<1.4$) and high-$z$ ($z>1.4$) GRBs, respectively. The GRB sample is taken from \citet{Liu:2014}. This sample consists of 59 low-$z$ GRBs and 79 high-$z$ GRBs. The rest of the paper is arranged as follows: In section \ref{sec:data}, we introduce the data and methodology that are necessary to our studies. We present our results in section \ref{sec:results}. Finally, discussions and conclusions are given in section \ref{sec:conclusions}.

\section{Data and methodology}\label{sec:data}

\subsection{Union2.1 and Hubble diagram}\label{sec:union2.1}

SNe Ia are usually regarded as ideal distance indicators to trace the Hubble diagram of the Universe due to their consistent luminosity.  The GRB candle, however, is much less standard than the SN Ia candle, since the explosion mechanism of GRBs is still not clearly known. Therefore, SNe Ia are often used to calibrate the distance moduli of GRBs. In this paper, we first use the recently published Union2.1 \citep{Suzuki:2012} dataset to constrain the cosmological parameters. The Union 2.1 dataset is a compilation of 580 well-observed SNe Ia in the redshift range $z\in[0.015,1.415]$. All the SNe Ia have high-quality light curves, so their distance moduli can be extracted with high precision. The Hubble diagram can be tightly constrained by the Union2.1 dataset.

In the spatially-flat isotropic spacetime, the luminosity distance can be expressed as a function of redshift as
\begin{equation}\label{lumi_distance}
d_L(z)=(1+z)\frac{c}{H_0}\int_0^{z} \frac{dz}{E(z)},
\end{equation}
where $c=3\times 10^8~{\rm m~s}^{-1}$ is the speed of light, $H_0=70~\rm{km}~\rm{s}^{-1}~\rm{Mpc}^{-1}$ is the Hubble constant at present time, and $E(z)\equiv H(z)/H_0$ is the normalized Hubble parameter. In the $\Lambda$CDM model, we have
\begin{equation}
E^2(z)=\Omega_M(1+z)^3+(1-\Omega_M),
\end{equation}
where $\Omega_M$ is matter density today. In the $w$CDM model, we have
\begin{equation}
E^2(z)=\Omega_M(1+z)^3+(1-\Omega_M)(1+z)^{3(1+w)},
\end{equation}
where $w=p/\rho$ denotes the equation of state of dark energy. In the Chevallier-Polarski-Linder (CPL) parametrization \citep{Chevallier:2001,Linder:2003}, the equation of state of dark energy is given as $w_{\rm de}=w_0+w_1z/(1+z)$, and in this case, $E(z)$ can be expressed as \citep{Nesseris:2004,Lazkoz:2005}
\begin{equation}
E^2(z)=\Omega_M(1+z)^3+(1-\Omega_M)(1+z)^{3(1+w_0+w_1)}\exp{\left(-\frac{3w_1z}{1+z}\right)}.
\end{equation}

In practice, it is convenient to define a dimensionless quantity called distance modulus, that is,
\begin{equation}\label{eq:dis_modulus}
\mu(z)=5\log\frac{d_L(z)}{\rm{Mpc}}+25,
\end{equation}
where ``$\log$" is the logarithm of base 10. The best-fit cosmological parameters can be derived by minimize $\chi^2$, i.e.,
\begin{equation}\label{eq:chi2}
  \chi^2=\sum_{i=1}^N\frac{(\mu^i_{\rm th}-\mu^i_{\rm obs})^2}{\sigma_{\mu_i}^2},
\end{equation}
where $\mu_{\rm th}$ is the theoretical distance modulus calculated from Eq.(\ref{eq:dis_modulus}), $\mu_{\rm obs}$ is the observed distance modulus, and $\sigma_{\mu}$ is the measurement error. $N=580$ is the number of SNe Ia in the Union2.1 dataset.

\subsection{GRBs and Amati relation}\label{sec:GRBs}

The Amati relation is a correlation between isotropic equivalent energy $E_{\rm iso}$ and spectrum peak energy in the comoving frame $E_{p,i}$ \citep{Amati:2002,Amati:2003,Amati:2006}. It was first discovered by \citet{Amati:2002} in 12 BeppoSAX GRBs of low redshift ($z\lesssim 2$, except one GRB), and was confirmed later in larger samples \citep{Amati:2003,Amati:2006}. The Amati relation can be parameterized as
\begin{equation}\label{eq:amati1}
  \log\frac{E_{\rm iso}}{{\rm erg}}=a+b\log\frac{E_{p,i}}{300~{\rm keV}},
\end{equation}
where
\begin{equation}\label{eq:iso_energy}
  E_{\rm iso}=4\pi d_L^2S_{\rm bolo}(1+z)^{-1}
\end{equation}
is the isotropic equivalent energy in the $1~\rm{keV}-10~\rm{MeV}$ energy band, and $S_{\rm bolo}$ is the bolometric fluence. The uncertainty of $E_{\rm iso}$ propagates from that of $S_{\rm bolo}$, i.e,
\begin{equation}
  \sigma_{E_{\rm iso}}=4\pi d_L^2\sigma_{S_{\rm bolo}}(1+z)^{-1}.
\end{equation}
The uncertainty from $d_L$ is absorbed into the intrinsic scatter $\sigma_{\rm int}$. Defining
\begin{equation}
  y\equiv\log\frac{E_{\rm iso}}{\rm erg},\quad  x\equiv\log\frac{E_{p,i}}{300~{\rm keV}},
\end{equation}
we can rewrite Eq.(\ref{eq:amati1}) as
\begin{equation}\label{eq:amati2}
  y=a+bx.
\end{equation}
The uncertainties of $y$ and $x$ are given as
\begin{equation}
  \sigma_y=\frac{1}{\ln 10}\frac{\sigma_{E_{\rm iso}}}{E_{\rm iso}},\quad  \sigma_x=\frac{1}{\ln 10}\frac{\sigma_{E_{p,i}}}{E_{p,i}},
\end{equation}
where ``$\ln$" represents the natural logarithm.

The slope and intercept of Amati relation, i.e., $b$ and $a$ in Eq.(\ref{eq:amati2}), can be obtained by directly fitting Eq.(\ref{eq:amati2}) to the observed GRB data. However, the plot of the Amati relation in the $(x,y)$ plane shows significant error bars in both the horizontal and vertical axes. Besides, the intrinsic scatter dominates over the measurement errors. Therefore, the ordinary least-$\chi^2$ method does not work. We may get different best-fit parameters depending on whether we minimize the sum of squared residuals in the $y$ axis or that in the $x$ axis. To avoid such a problem, we use the fitting method presented in \citet{DAgostini:2005}. The joint likelihood function for the slope $b$, intercept $a$, and intrinsic scatter $\sigma_{\rm int}$ is given as
\begin{equation}\label{eq:likelihood}
  \mathcal{L}(\sigma_{\rm int},a,b)\propto\prod_i\frac{1}{\sqrt{\sigma_{\rm int}^2+\sigma_{y_i}^2+b^2\sigma_{x_i}^2}}
  \times \exp\left[-\frac{(y_i-a-bx_i)^2}{2(\sigma_{\rm int}^2+\sigma_{y_i}^2+b^2\sigma_{x_i}^2)}\right].
\end{equation}
The minus-log-likelihood is given as
\begin{equation}\label{eq:loglikelihood}
  -\ln\mathcal{L}(\sigma_{\rm int},a,b)=\frac{1}{2}\sum_i\ln(\sigma_{\rm int}^2+\sigma_{y_i}^2+b^2\sigma_{x_i}^2) +\frac{1}{2}\sum_i\frac{(y_i-a-bx_i)^2}{\sigma_{\rm int}^2+\sigma_{y_i}^2+b^2\sigma_{x_i}^2}+{\rm const}.
\end{equation}
The sum runs over all the GRB data points. The best-fit parameters are the ones which can minimize the right-hand-side of Eq.(\ref{eq:loglikelihood}). Note that the Amati relation depends on the luminosity distance $d_L$, which further depends on the specific cosmological model.

The GRB dataset used to our analysis is directly taken from \citet{Liu:2014}. The dataset is a simple summation of samples presented in two previously published papers \citep{Wei:2010,Qin:2013}. The two papers collected the data from many other published papers. This dataset is at present the largest and  most up-to-date sample available to analyze the Amati relation. This sample is a collection of GRBs observed by various instruments, such as BATSE, BeppoSAX, HETE-2, Konus-Wind, Swift, Fermi, and so on. The spectra properties (photon indices, peak energy, fluence, etc.) of these GRBs are measured with enough precision. The redshifts are well determined through the observation of afterglows. The bolometric fluence is calculated in the rest frame $1-10,000$ keV energy band by using the Band function (Schaefer 2007). Our sample in total includes 59 low-$z$ ($z<1.4$) and 79 high-$z$ ($z>1.4$) GRBs. We analysis the low-$z$ and high-$z$ GRBs separately. To check the possible model-dependence of the Amati relation, we investigate it in three different cosmological models, i.e., $\Lambda$CDM model, $w$CDM model and  CPL model.

\section{Results}\label{sec:results}

To test the Amati relation, the Hubble diagram of the Universe should be known a priori. Thus, we first use the Union2.1 dataset to constrain the Hubble diagram of the Universe in low-redshift range ($z<1.4$). The constraints on the parameters of three cosmological models are listed in Table \ref{tab:cosmol_parameters}.
\begin{table}
\centering
\caption{\small{The best-fit cosmological parameters and their $1\sigma$ uncertainties from the Union2.1 dataset in three different cosmological models.}}\label{tab:cosmol_parameters}
\begin{tabular}{c|cccc}
  \hline\hline
    & $\Omega_M$ & $w$ & $w_0$ & $w_1$ \\
  \hline
  $\Lambda$CDM & $0.2798\pm 0.0130$ & -- & -- & -- \\
  $w$CDM & $0.2755\pm 0.0640$ & $-0.9903\pm 0.1431$ & -- & -- \\
  CPL & $0.2962\pm 0.2224$ & -- & $-1.0090\pm 0.2249$ & $-0.2455\pm 2.9514$ \\
  \hline
\end{tabular}
\end{table}
The quoted errors are of $1\sigma$. In the fitting procedure, we fix the Hubble constant to be $H_0=70.0~{\rm km~s}^{-1}~{\rm Mpc}^{-1}$. Note that the parameters of $\Lambda$CDM model and $w$CDM model can be tightly constrained. However, the constraint on the CPL model is rather loose, especially on the parameter $w_1$.

\begin{figure}
  \centering
 \includegraphics[width=16 cm]{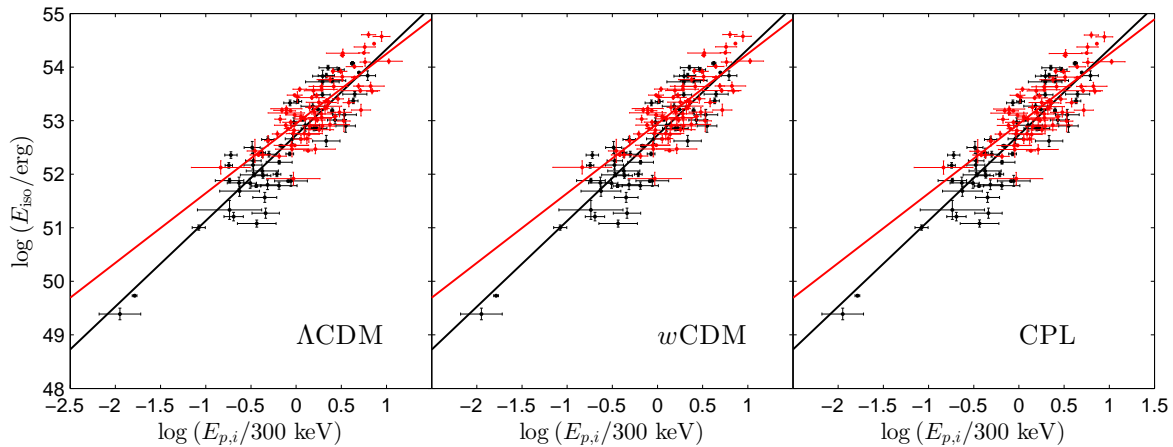}
 \caption{\small{The Amati relation for low-$z$ (black line) and high-$z$ (red line) GRBs in different cosmological models. From left to right: $\Lambda$CDM model, $w$CDM model and CPL model.}}\label{fig:amati}
\end{figure}

\begin{figure}
  \centering
 \includegraphics[width=16 cm]{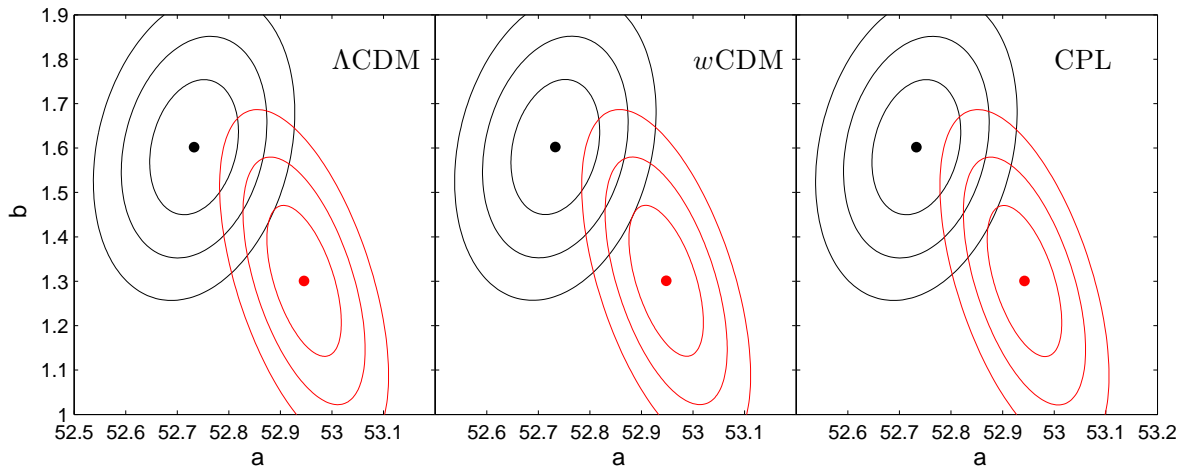}
 \caption{\small{The $1\sigma$, $2\sigma$ and $3\sigma$ contours in the $(a,b)$ plane for low-$z$ (black curves) and high-$z$ (red curves) GRBs in different cosmological models. The central values are denoted by dots. From left to right: $\Lambda$CDM model, $w$CDM model and CPL model.}}\label{fig:contour}
\end{figure}

We directly extrapolate the Hubble diagram to the whole redshift range. Assuming that GRBs follow the same Hubble diagram, we can calculate the luminosity distance of GRBs from Eq.(\ref{lumi_distance}), as well as the isotropic equivalent energy from Eq.(\ref{eq:iso_energy}). The Amati relation Eq.(\ref{eq:amati2}) is then used to fit the low-$z$ and high-$z$ GRBs, respectively. The best-fit parameters are listed in Table \ref{tab:amati_parameters}. The quoted errors are of $1\sigma$.
\begin{table}
\centering
\caption{\small{The intrinsic scatters, intercepts and slopes of Amati relation for low-$z$ and high-$z$ GRBs in three different cosmological models.}}\label{tab:amati_parameters}
\begin{tabular}{ll|ccc}
  \hline\hline
    &   & $\sigma_{\rm int}$ & $a$ & $b$ \\
  \hline
  $\Lambda$CDM:  & low-$z$ & $0.3810\pm 0.0433$ & $52.7326\pm 0.0568$ & $1.6020\pm 0.1004$ \\
               & high-$z$ & $0.3133\pm 0.0313$ & $52.9459\pm 0.0476$ & $1.3008\pm 0.1122$ \\
  \hline
  $w$CDM:       & low-$z$ & $0.3810\pm 0.0433$ & $52.7330\pm 0.0568$ & $1.6023\pm 0.1004$ \\
               & high-$z$ & $0.3132\pm 0.0313$ & $52.9479\pm 0.0476$ & $1.3010\pm 0.1121$ \\
  \hline
  CPL:           & low-$z$ & $0.3810\pm 0.0433$ & $52.7328\pm 0.0568$ & $1.6020\pm 0.1004$ \\
               & high-$z$ & $0.3135\pm 0.0313$ & $52.9423\pm 0.0476$ & $1.3005\pm 0.1122$ \\
  \hline
\end{tabular}
\end{table}
The $E_{\rm peak}-E_{\rm iso}$ correlation is plotted in Figure \ref{fig:amati} in logarithmic coordinates. Low-$z$ and high-$z$ GRBs are denoted by black and red dots with $1\sigma$ error bars, respectively. The lines stand for the best-fit results.
Besides, the $1\sigma$, $2\sigma$ and $3\sigma$ contours in the $(a,b)$ plane for low-$z$ (black curves) and high-$z$ (red curves) GRBs are plotted in Figure \ref{fig:contour}. The best-fit central values are denoted by dots.

From Table \ref{tab:amati_parameters} and Figure \ref{fig:amati}, we can see that the slope and intercept of low-$z$ GRBs differ from that of high-$z$ GRBs significantly. Low-$z$ GRBs have larger slope, but smaller intercept than high-$z$ GRBs. This can be seen more clearly from the contour plot in the ($a,b$) plane in Figure \ref{fig:contour}. There is no overlap between $1\sigma$ contours, while only a little overlap between $2\sigma$ contours. The best-fit central values of low-$z$ GRBs do not locate in the $3\sigma$ contour of high-$z$ GRBs. Similarly, the best-fit central values of high-$z$ GRBs locate outside of the $3\sigma$ contour of low-$z$ GRBs. This means that the Amati relation of low-$z$ GRBs differs from that of high-$z$ GRBs at more than $3\sigma$ confidence level. Another interesting feature is that, there seems to be a positive correlation between the slope and intercept for low-$z$ GRBs. While for high-$z$ GRBs, the slope is negatively correlated to the intercept. We can also see that the Amati relation is insensitive to cosmological models.

We note that most high-$z$ GRBs are energetic ($E_{\rm iso}>10^{52}$ ergs) and have large peak energy ($E_{p,i}>300$ keV). On the contrary, low-$z$ GRBs are in general less energetic and have smaller $E_{p,i}$ than high-$z$ GRBs. Especially, there are two dim GRBs with isotropic equivalent energy $E_{\rm iso}<10^{50}$, while most other low-$z$ GRBs have $E_{\rm iso}>10^{51}$. The best fit line for low-$z$ GRBs is largely affected by the two dim GRBs. GRBs with isotropic equivalent energy small than $10^{51}$ ergs tend to be detected only at low redshift, so that the $E_{\rm iso}$ coverage in redshift may be biased. This may be a reason why the Amati relations for low-$z$ and high-$z$ GRBs have very different slopes.

\section{Discussions and conclusions}\label{sec:conclusions}

In this paper, we investigated the Amati relation for low-$z$ and high-$z$ GRBs separately in three cosmological models, i.e., $\Lambda$CDM model, $w$CDM model and CPL model. SNe Ia were used to constrain the parameters of background cosmos. It was found that the Amati relation of low-$z$ GRBs differs from that of high-$z$ GRBs at more than $3\sigma$ confidence level. We noted that the slope of low-$z$ GRB is larger than that of high-$z$ GRBs, while high-$z$ GRBs have larger intercept than low-$z$ GRBs. Our results are consistent with that of \citet{Wang:2011}. We also found a positive correlation between the slope and intercept for low-$z$ GRBs. While for high-$z$ GRBs, the correlation is negative. These features do not depend on the cosmological models. We noted that high-$z$ GRBs are often much energetic than low-$z$ GRBs. The $E_{\rm iso}$ coverage in redshift may be biased since dim GRBs are only detected at low redshift.

The Amati relation is often used to calibrate the distance moduli of GRBs. These calibrating methods are based on the assumption that the Amati relation is universal over all redshifts. The redshift-dependence of Amati relation, as indicated in this paper, puts a challenge on this assumption. However, this does not necessarily  mean that the Amati relation indeed varies with redshift. Another possibility is that the high-$z$ Hubble diagram deviates from the low-$z$ Hubble diagram predicted by SNe Ia. But this is less likely, because the constraints on the Hubble diagram from the combination of SNe Ia and GRBs (which are calibrated using the Amati relation) is very close to that from SNe Ia alone \citep{Liu:2014}.

The Amati relation has intrinsic scatter much larger than the measurement errors. The uncertainties of distance moduli of GRBs calibrated through Amati relation is about one order of magnitude larger than the uncertainties of SNe Ia \citep{Liu:2014}. When combining GRBs with SNe Ia to constrain cosmological parameters, the weight of GRBs is about two orders of magnitude smaller than that of SNe Ia, since the weight of a data point is inversely proportional to the square of its uncertainty (see Eq.(\ref{eq:chi2})). This is one reason why adding GRBs to the Union2.1 dataset (or any other datasets which have much higher precision than GRBs) has only small effect on constraining the cosmological parameters \citep{Liu:2014}. Another reason is that the number of GRBs is much smaller than the number of SNe Ia. In addition, use GRBs alone to constrain the cosmological parameters will lead to unreasonable results. All of these arouse us to search for other calibrating methods. For example, \citet{Basak:2013} pointed out that the pulse-wise Amati relation is more robust than the time-averaged one. The uncertainties should be much reduced before GRBs can be used, in combination with other candles, to trace the Hubble diagram of the Universe.

\section*{Acknowledgements}
We are grateful to Y. Sang, P. Wang and D. Zhao for useful discussions. This work has been funded by the National Natural Science Fund of China under grants Nos. 11375203, 11305181, 11322545 and 11335012, and by the project of Knowledge Innovation Program of Chinese Academy of Sciences.

\label{lastpage}

\end{document}